\documentclass[preprint]{iucr}

\usepackage{graphicx}
\usepackage[utf8]{inputenc}
\usepackage{url}

     \paperprodcode{a000000}      
     \paperref{xx9999}            
     \papertype{FA}               

     \paperlang{english}          
     \journalcode{J}              
     \journalyr{2009}
     \journaliss{1}
     \journalvol{42}
     \journalfirstpage{000}
     \journallastpage{000}
     \journalreceived{0 XXXXXXX 0000}
     \journalaccepted{0 XXXXXXX 0000}
     \journalonline{0 XXXXXXX 0000}

\begin{document}                  



\title{Crystal structure solution from experimentally determined atomic
pair distribution functions}


\cauthor[a]{P.}{Juhás}{pj2192@columbia.edu}{}
\author[b]{L.}{Granlund}
\author[b]{S.~R.}{Gujarathi}
\author[b]{P.~M.}{Duxbury}
\author[a,c]{S.~J.~L.}{Billinge}

\aff[a]{{Department of Applied Physics and Applied Mathematics,
    Columbia University},
    \city{{New York}, New York, 10027, \country{USA}}}
\aff[b]{{Department of Physics and Astronomy,
    Michigan State University},
    \city{{East Lansing}, Michigan, 48824, \country{USA}}}
\aff[c]{{Condensed Matter Physics and Materials Science Department,
    Brookhaven National Laboratory},
    \city{Upton, New York, 11973, \country{USA}}}

\maketitle                        


\begin{abstract}
The paper describes an extension of the Liga algorithm for structure
solution from atomic pair distribution function (PDF), to handle periodic
crystal structures with multiple elements in the unit cell.
The procedure is performed in 2 separate steps - at first the
Liga algorithm is used to find unit cell sites consistent with
pair distances extracted from the experimental PDF.  In the second step the
assignment of atom species over cell sites is solved by minimizing the
overlap of their empirical atomic radii.  The procedure has been
demonstrated on synchrotron x-ray PDF data from 16 test samples.
The structure solution was successful for 14 samples including cases
with enlarged super cells.  The algorithm success rate and the reasons
for failed cases are discussed together with enhancements that should
improve its convergence and usability.
\end{abstract}


\section{Introduction}
\label{Introduction}
Crystallographic methods of structure solution are the gold-standard for determining atomic arrangements in crystals including, in the absence of single crystals, structure solution from powder diffraction data~\cite{pecha;b;fopdascom05,david;b;sdfpdd02}.  Here we show that crystal structure solution is also possible from experimentally determined atomic pair distribution functions (PDF) using the Liga algorithm that was developed for nanostructure determination~\cite{juhas;n06,juhas;aca08}.  The PDF is the Fourier transform of the properly normalized intensity data from an isotropically scattering sample such as a glass or a crystalline powder.  It is increasingly used as a powerful way to study atomic structure in nanostructured materials~\cite{billi;jssc08,egami;b;utbp03}.  Such nanostructures do not scatter with well defined Bragg peaks and are not amenable to crystallographic analysis~\cite{billi;s07}, but refinements of models to PDF data yield quantitatively reliable structural information~\cite{proff;jac99,farro;jpcm07,tucke;jpcm07}.  Recently \emph{ab initio} structure solution was demonstrated from PDF data of small elemental clusters~\cite{juhas;n06}.  Here we show that these methods can be extended to solve the structure of a range of crystalline materials.

Whilst it is unlikely that this kind of structure solution will replace crystallographic methods for well ordered crystals, this work demonstrates both that structure solution from PDF data can be extended to compounds, and that robust structure solutions are possible from the experimentally determined PDFs of a wide range of materials.  We also note that there may be an application for this approach when the space-group of the crystal is not known, as the Liga algorithm does not make use of such symmetry information.  In fact, the space group can be determined afterwards analyzing the symmetry of the solved electron density map \cite{palat;jac08}. 
However, this approach is promising for the case where the local
structure deviates from the average crystallographic structure, as has
been observed in a number of complex crystals, for example the
magnetoresistive La$_{1-x}$Ca$_x$MnO$_3$ system~\cite{qiu;prl05,bozin;pb06} or ferroelectric lead-based perovskites~\cite{dmows;jpcs00,juhas;prb04}.
The PDF contains this local information due to the inclusion of diffuse scattering intensities in the Fourier transform and it is possible to focus the modeling on a
specific length-scale when searching for matching structure models, allowing in principle structure solutions of local, intermediate, and long-range order to be obtained separately.

The procedure assumes a
periodic system with known lattice parameters and stoichiometry,
otherwise there is no information on location or symmetry of the
atom sites in the unit cell.  To solve the unit cell structure the technique
constructs a series of trial clusters using the PDF-extracted
distances.  The tested structures are created with a direct use of
distance information in the experimental data giving it significantly better performance than procedures that search by
random structure updates such as Monte Carlo based minimization schemes~\cite{juhas;n06,juhas;aca08}.

\section{Experimental procedures}
\label{ExperimentalProcedures}

The extended Liga procedure has been tested with
experimental x-ray PDFs collected from inorganic test materials.
Powder samples of Ag, BaTiO$_{3}$, C-graphite, CaTiO$_3$, CdSe,
CeO$_2$, NaCl, Ni, PbS, PbTe, Si, SrTiO$_3$, TiO$_2$ (rutile),
Zn, ZnS (sphalerite) and ZnS (wurtzite) were obtained from commercial
suppliers.  Samples were ground in agate mortar to decrease their
crystallite size and improve powder averaging.
The experimental PDFs were measured using synchrotron
x-ray diffraction at the 6ID-D beamline of the Advanced Photon Source,
Argonne National Laboratory using the x-ray energies of 87 and 98~keV.
The samples were mounted using a thin kapton tape in a circular,
10~mm hole of a 1~mm thick flat plate holder, which was positioned
in transmission geometry with respect to the beam.  The x-ray data were
measured using the ``Rapid Acquisition'' (RA-PDF) setup, where the
diffracted intensities were scanned by a MAR345 image plate
detector, placed about 20~cm behind the sample~\cite{chupa;jac03}.
All measurements were performed at room temperature.
The raw detector images were integrated using the Fit2D program
\cite{hamme;esrf98} to reduce them to a standard intensity
vs.\ $2\theta$ powder data.  The integrated data were then
converted by the PDFgetX2 program~\cite{qiu;jac04i} to experimental PDFs.
The conversion to PDF was conducted with corrections for Compton
scattering, polarization and fluorescence effect, as available in
the PDFgetX2 program.  The maximum value of the scattering
wavevector $Q_{\max}$ ranged from 19~Å$^{-1}$ to 29~Å$^{-1}$,
based on a visual inspection of the noise in the $F(Q) = Q [S(Q) - 1]$
curves.

The PDF function $G(r)$ was obtained by a Fourier transformation
of $F(Q)$,
\begin{equation}
\label{eq;sqtogr}
G(r) = \frac{2}{\pi}\int_{Q_{\min}}^{Q_{\max}} F(Q) \sin Qr \> \mathrm{d}Q,
\end{equation}
and provided a scaled measure of finding a pair of atoms separated by
distance $r$
\begin{equation}
\label{eq;grassum}
G(r) = \frac{1}{N r \langle f \rangle^2}
       \sum_{i \neq j} f_i f_j \delta(r - r_{ij})  -  4 \pi \rho_0 r.
\end{equation}
The $G(r)$ function has a convenient property that its peak amplitudes
and standard deviations remain essentially constant with $r$ and is thus
suitable for curve fitting.  A detailed discussions of the PDF theory,
data acquisition and applications for structure analysis can be found
in~\cite{egami;b;utbp03,farro;aca09}.

\section{Methods}
\label{Methods}

The structure solution procedure was carried out in three
separate steps, as described in the sections below.
The first step consists of peak search and profile fitting in
the experimental PDF to identify prominent inter-atomic distances up to
a cutoff distance $d_{\mathit{cut}}$.
We have developed an automated
peak extraction method which eases this task.  In the second step these
distances are used as inputs for the Liga algorithm, which searches for
unit cell positions that give structure with the best match in pair
lengths.  If the sample has several chemical species, a final ``coloring'' step
is necessary to assign proper atom species to the unit cell
sites.  This can be done by making use of PDF peak amplitude information.
However, we have found that coloring can be also solved by optimizing the
overlap of the empirical atom radii at the neighboring sites, which is
simpler to implement and works with greater reliability.

To verify the quality and uniqueness of the structure, the Liga
algorithm has been run for each sample multiple (at least 10) times
with the same inputs, but different seeds of the random number generator.  For most
samples the resulting structures were all equivalent, but sometimes
the program gave several geometries with similar agreement to
the PDF-extracted pair distances.  In all these cases the correct
structure could be resolved in the coloring step, where it displayed
significantly lower atom radii overlap and converged to known structure
solution.  A small number of structures would not solve by this process
and the reasons for failure are discussed below.

\subsection{Extraction of pair distances from the experimental PDF}
In the PDF frequent
pair distances generate sharp peaks in the measured $G(r)$ curve
with amplitudes following Equation~(\ref{eq;grassum}).  The peaks
are broadened
to approximately Gaussian shape that reflects atom thermal vibrations
and limited experimental resolution.  Additional broadening and
oscillations are introduced to the PDF due to the maximum wavevector
$Q_{\max}$ that can be achieved in the measurement.  This cutoff in
$Q_{\max}$ in effect convolutes ideal peak profiles with a
sinc function $\sin(Q_{\max} r) / r$ thus creating satellite
termination ripples.

Recovering the underlying peaks from the PDF is not trivial.  The
experimental curve can have false peaks due to termination ripples.
Nearby peaks can overlap and produce complicated profiles that are
difficult to decompose.  To simplify the process of extracting
inter-atomic distances we have developed an automated method for peak
fitting that adds peak profiles to fit the data to some user-defined
tolerance, while using as few peaks as possible to avoid over-fitting.  This
method grows peak-like clusters of data points while fitting one or more
model peaks to each cluster.  Adjacent clusters iteratively combine
until there is a single cluster with a model that fits the entire data
set.  This allows a steady growth in model complexity by progressively
refining earlier and less accurate models.  Furthermore, most adjustable
parameters can be estimated, in principle, from experimental knowns.  A
full description of the peak extraction method will be presented in a
future paper.

The present work uses the simplest model for peaks, fitting the $G(r)$
data with Gaussian peaks over $r$ and using an assumed value of $\rho_{0}$.  This
model ignores the effect of termination ripples, but for our data the
spurious peaks due to these ripples were usually identifiable by
their small size.  Furthermore, the Liga algorithm is not required to
use every distance it is given, and should exhibit a limited tolerance
of faulty distances.  The peak fitting procedure returns positions,
widths and integrated areas of the extracted peaks, of which only the
peak positions were used for structure determination.

The peak extraction procedure was implemented in Mathematica 6
and tested on both the experimental and
simulated data.  A typical runtime was about 5 minutes.
Since the structures are known we can compare the results of the peak extraction with
the expected results.  For both experimental and simulated PDFs of the tested structures
these compared qualitatively well to the ideal distances up to
$\sim$10-15~Å, including accurate identification of some
obscured peaks.  Past that range the number of distinct, but very close,
distances in the actual structure is so great that reliable peak
extraction is much more difficult.  For this reason we only performed
peak extraction up to 10~Å before running the trials
described in Section~\ref{Results}.  Apart from removing peaks below a noise
threshold in order to filter termination ripples out, and one difficult
peak in the graphite data, all distances used in the structure solution
trials below come directly from the peak extraction method.

\subsection{Unit cell reconstruction using the Liga algorithm}

\begin{figure}
\includegraphics[clip=true]{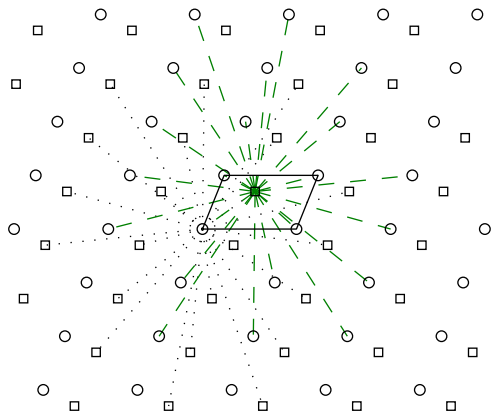}
\label{fig2dLattice}
\caption{Schematic calculation of the distance cost $C_d$.  To achieve
low $C_d$ a unit cell site needs to be at a correct distance from
other cell sites and from their translational images.
}
\end{figure}

In the second step the Liga algorithm searches for the atom
positions in the unit cell that make the best agreement
to the extracted pair distances.  The quality of distance match
is expressed by cost $C_d$ defined as a mean square difference between
observed and modeled pair distances.
\begin{equation}
\label{eq;ligacost}
C_d = \frac{1}{P} \sum_{d_k < d_{\mathit{cut}}}
    \left( t_{k,\mathit{near}} - d_k \right)^2
\end{equation}
The index $k$ goes over all pair distances $d_k$ in the model that are
shorter than the cutoff length $d_{\mathit{cut}}$ and compares them with
the nearest observed distance $t_{k,\mathit{near}}$, while $P$ is the
number of model distances.  This cost definition considers
only distance values as extracted from the PDF peak positions,
and ignores their relative occurrences.
For multi-component systems there is in fact no
straightforward way of extracting distance multiplicities,
because it is not known what atom pairs are present in each PDF peak.
Nevertheless, the cost definition still imposes strict
requirements on the model structure, as displayed in
Fig.~\ref{fig2dLattice}.  A
site in the unit cell must be at a good, matching distance not
only from all other cell sites, but also from all of their translational
images within the cutoff radius.

To find an optimum atom position in the unit cell structure the
Liga algorithm uses input pair distances in an iterative build-up and
disassembly of partial structures~\cite{juhas;n06}.
The procedure maintains a pool of partial unit cell structures at each
possible size from a single atom up to a complete unit cell.
These ``candidate clusters'' are assigned to ``divisions''
according to the number of sites they contain, therefore there are
as many divisions as is the number of atoms in a complete unit cell.
The structures at each division compete
against each other in a stochastic process, where the probability of
winning equals the reciprocal distance cost $C_d$, and a win is thus more
likely for low-cost structures.  The winning cluster is selected for
``promotion,''
where it adds one or more atoms to the
structure and thus advances to a higher division.  At the new division a
poorly performing high-cost candidate is ``relegated'' to the original
division of the promoted structure, thus keeping the total number of
structures at each division constant.  The relegation is accomplished
by removing cell sites that have the largest contributions to the
total cost of the structure.  Both promotion and relegation
steps are followed by downhill relaxation of the worst site,
i.e., the site with the largest share of the total cost $C_d$.
The process of promotion and relegation is performed at every division
in a ``season'' of competitions.  These seasons are repeated many
times until a full sized structure
attains sufficiently low cost or until a user-specified time
limit.  A complete description of the Liga algorithm details
can be found in~\cite{juhas;aca08}.

\subsection{Atom assignment}

The Liga algorithm used in the structure solution step has no notion of
chemical species and therefore returns only coordinates of the atom
sites in the unit cell.  For a multi-component system an
additional step, dubbed coloring, is necessary to assign
chemical elements over known cell sites.  To assess the
quality of different assignments we have tested two definitions
for a cost of a particular coloring.  The first method uses
a weighted residuum, $R_w$, from a least-squares PDF
refinement to the input PDF data~\cite{egami;b;utbp03}.
The PDF refinement was performed with a fully automated PDFfit2 script,
where the atom positions were all fixed and only the atomic displacement
factors, PDF scale factor and $Q$-resolution damping factor were
allowed to vary.  The second procedure
defines coloring cost $C_c$ as an average overlap of the empirical
atomic radii, so that
\begin{equation}
\label{eq;coloringcost}
C_c = \frac{1}{N} \sum_{d_{k} < r_{k,1} + r_{k,2}}
    \left( r_{k,1} + r_{k,2} - d_{k} \right)^2
\end{equation}
The index $k$ runs over all atom pairs considering periodic boundary
conditions,
$r_{k,1}$ and $r_{k,2}$ are the empirical radii values of
the first and second atom in the pair $k$, and $N$ is the number of
atoms in the unit cell.

Considering an $N$ atom structure with $s$ different atom species,
the total number of possible assignments is given by the multinomial
expression $N! / (n_1! \: n_2! \: \ldots \: n_s!)$\@.
For a 1:1 binary system the number of possible assignments tends to
$2^N$ with increasing $N$.  Such exponential growth in possible
configurations makes it quickly impossible to compare them all in
an exhaustive way.
We have therefore employed a simple downhill search, which starts with
a random element assignment.  The initial coloring cost $C_c$ is calculated
together with a cost change for every possible
swap of two atoms between unit cell sites.  The site flip that
results in the largest decrease of the total coloring cost is accepted and all
cost differences are evaluated again.  The site swap is then repeated
until a minimum configuration is achieved, where all site flips
increase the coloring cost.  The downhill procedure was
verified by repeating it 5 times using different initial assignments.
In nearly all cases these runs converged to the same
atom configurations.

The downhill procedure was performed using both definitions of the
coloring cost.  For the coloring cost obtained by PDF fitting the
procedure was an order of magnitude slower and less reliable, as the
underlying PDF refinements could converge badly for poor atom assignments.
The second method, which calculated cost from radii-overlap, was
considerably
faster and more robust.  For all tested materials, the overlap-based
coloring assigned all atoms correctly, when run on correct structure
geometry.  The overlap cost was evaluated using either the covalent
radii by~\cite{corde;d08}
or the ionic radii from~\cite{shann;aca76} for more ionic
compounds.  For some ions the Shannon table provides
several radii values depending on their coordination number or
spin state.  Although these variants in ionic radii can vary by
as much as about 30\%, the choice of particular radius had no
effect on the best assignment for all studied structures.

\section{Results}
\label{Results}

The experimental x-ray PDFs were acquired from 16 test samples
with well known crystal structures.  To verify that the measured
PDF data were consistent with established structure results structure
refinements of the known structures were carried out using the PDFgui program~\cite{farro;jpcm07}.  The PDF fits were done with structure data
obtained from the Crystallography Open Database (COD)~\cite{grazu;jac09}.
The structure parameters were all kept constant in the refinements,
which modified only parameters related to the PDF extraction, such
as PDF scale, $Q$ resolution dampening envelope and a small rescaling
of the lattice parameters.  These refinements are summarized in
Table~\ref{tab;PDFrefinements}, where low values of the fitting residual
$R_w$ confirm good agreement between experimental PDFs and expected
structure results.

The PDF datasets were then subjected to the peak search, Liga structure
solution and coloring procedures as described above.  To check the
stability of this method, several structures were solved using an enlarged
periodicity of 1$\times$1$\times$2, 1$\times$2$\times$2 or
2$\times$2$\times$2 super cells.  The lattice parameters used in the
Liga crystallography step were obtained from the positions of the
nearest PDF peaks.  In several cases, such as for BaTiO$_{3}$ where peak
search could not resolve tetragonal splitting, the cell parameters were
taken from the respective CIF reference, as listed in
Table~\ref{tab;PDFrefinements}.

The structure solution was considered successful if the found structure
displayed the same nearest neighbor coordination as its CIF reference and
no site was offset by more than 0.3~Å from its correct
position.  The solution accuracy was evaluated by finding the best
overlay of the found structure to the reference CIF data.  The
optimum overlay was obtained by an exhaustive search over all symmetry
operations defined in the CIF file and over all mappings of solved atom
sites to all reference sites containing the same element.  The
overlaid structures were then compared for the differences in
fractional coordinates and for the root mean square distortion $s_r$ of the solved
sites from their correct positions.  Table~\ref{tab;SolvedStructures}
shows a summary of these results for all tested structures.

\begin{table}
\caption{List of measured x-ray PDFs and their fitting residua
$R_w$ with respect to established structures from the literature.
}
\label{tab;PDFrefinements}
\begin{tabular}{lll}
\hline
sample              & $R_w$ & CIF reference \\
\hline
Ag                  & 0.095 & \cite{wycko;bk63} \\
BaTiO$_{3}$         & 0.123 & \cite{megaw;ac62} \\
C (graphite)        & 0.248 & \cite{wycko;bk63} \\
CaTiO$_{3}$         & 0.083 & \cite{sasak;acc87} \\
CdSe                & 0.149 & \cite{wycko;bk63} \\
CeO$_{2}$           & 0.098 & \cite{wycko;bk63} \\
NaCl                & 0.161 & \cite{jurge;ic00} \\
Ni                  & 0.109 & \cite{wycko;bk63} \\
PbS                 & 0.085 & \cite{ramsd;ami25} \\
PbTe                & 0.070 & \cite{wycko;bk63} \\
Si                  & 0.085 & \cite{wycko;bk63} \\
SrTiO$_{3}$         & 0.143 & \cite{mitch;apa02} \\
TiO$_{2}$ (rutile)  & 0.146 & \cite{meagh;canmin79} \\
Zn                  & 0.105 & \cite{wycko;bk63} \\
ZnS (sphalerite)    & 0.102 & \cite{skinn;ami61} \\
ZnS (wurtzite)      & 0.174$^1$ & \cite{wycko;bk63} \\
\hline
\multicolumn{3}{l}{
$^1$ refined as mixture of wurtzite and sphalerite phases
} \\
\hline
\end{tabular}
\end{table}

The procedure converged to a correct structure for 14 out of 16 studied
samples and failed to find the remaining 2.  The convergence was
more robust for high-symmetry structures, such as Ag (\textit{f.c.c.}),
NaCl or ZnS sphalerite, which could be reliably solved also in
enlarged [222] supercells.  For all successful runs the distance cost $C_d$
of the Liga-solved structure was comparable to the one from the CIF
reference and the atom overlap measure $C_c$ was close to zero.
ZnS sphalerite shows a notable difference between the $C_d$ values of
the solution and its CIF reference, however this was caused by using
a PDF peak position as a cell parameter for the solved structure.
Apparently the PDF peak extracted at $r \approx a$ was slightly
offset with respect to other peaks, nevertheless the Liga algorithm
still produced atom sites with correct fractional coordinates.  The mean
displacement $s_r$ for ZnS is 0~Å, because solved structures and CIF
references were compared using lattice parameters rescaled to their
CIF values.

\begin{table}
\caption{Summary of tested structure solutions from x-ray PDF data}
\label{tab;SolvedStructures}
\begin{tabular}{lr*{8}{l}}
\hline
\multicolumn{2}{l}{sample \hfill atoms} &
    \multicolumn{2}{l}{cost $C_d$ (0.01~Å$^2$)} &
    \multicolumn{2}{l}{cost $C_c$ (Å$^2$)} &
    \multicolumn{4}{l}{deviation of coordinates} \\
(supercell) & & Liga & CIF & Liga & CIF &
    $s_x$ & $s_y$ & $s_z$ & $s_r$ (Å) \\
\hline
\multicolumn{10}{l}{successful solutions} \\
\hline
Ag [111] & 4 & 0.0232 & 0.136 & 0 & 0.001 &
    0 & 0 & 0 & 0 \\
Ag [222] & 32 & 0.0097 & 0.136 & 0 & 0.001 &
    0.00025 & 0.00024 & 0.00003 & 0.0014 \\
BaTiO$_3$ [111] & 5 & 0.370 & 0.394 & 0.040 & 0.042 &
    0.0057 & 0.0066 & 0.014 & 0.064 \\
BaTiO$_3$ [112] & 10 & 0.392 & 0.394 & 0.058 & 0.042 &
    0.00023 & 0.039 & 0.018 & 0.16 \\
C graphite [111] & 4 & 0.396 & 0.574 & 0.010 & 0.016 &
    0.0029 & 0.0029 & 0.036 & 0.14 \\
C graphite [221] & 16 & 0.420 & 0.574 & 0.010 & 0.016 &
    0.0086 & 0.0065 & 0.036 & 0.15 \\
CdSe [111] & 4 & 0.107 & 0.138 & 0 & 0.001 &
    0 & 0 & 0.0055 & 0.027 \\
CdSe [221] & 16 & 0.0856 & 0.138 & 0 & 0.001 &
    0.00010 & 0.00013 & 0.0057 & 0.028 \\
CeO$_2$ [111] & 12 & 0.515 & 0.554 & 0 & 0 &
    0 & 0 & 0 & 0 \\
NaCl [111] & 8 & 1.75 & 1.71 & 0 & 0 &
    0 & 0 & 0 & 0 \\
NaCl [222] & 64 & 1.20 & 1.71 & 0 & 0 &
    0.00031 & 0.00031 & 0.00035 & 0.0032 \\
Ni [111] & 4 & 0.0024 & 0.0024 & 0 & 0 &
    0 & 0 & 0 & 0 \\
Ni [222] & 32 & 0.0025 & 0.0024 & 0 & 0 &
    0.00015 & 0.00013 & 0.00013 & 0.0008 \\
PbS [111] & 8 & 0.0125 & 0.0104 & 0.010 & 0.011 &
    0 & 0 & 0 & 0 \\
PbS [222] & 64 & 0.0140 & 0.0104 & 0.010 & 0.011 &
    0.00005 & 0.00004 & 0.00005 & 0.0005 \\
PbTe [111] & 8 & 0.0024 & 0.0127 & 0.097 & 0.090 &
    0 & 0 & 0 & 0 \\
PbTe [222] & 64 & 0.0022 & 0.0127 & 0.097 & 0.090 &
    0.00011 & 0.00011 & 0.00008 & 0.0011 \\
Si [111] & 8 & 0.0045 & 0.0045 & 0 & 0 &
    0 & 0 & 0 & 0 \\
Si [222] & 64 & 0.0048 & 0.0045 & 0 & 0 &
    0.00010 & 0.00009 & 0.00008 & 0.0009 \\
SrTiO$_3$ [111] & 5 & 0.437 & 0.437 & 0.002 & 0.002 &
    0 & 0 & 0 & 0 \\
Zn [111] & 2 & 0.495 & 0.470 & 0 & 0 &
    0 & 0 & 0.027 & 0.095 \\
Zn [222] & 16 & 0.564 & 0.470 & 0 & 0 &
    0.00010 & 0.00006 & 0.020 & 0.080 \\
ZnS sphalerite [111] & 8 & 0.150 & 0.0647 & 0 & 0 &
    0 & 0 & 0 & 0 \\
ZnS sphalerite [222] & 64 & 0.160 & 0.0647 & 0 & 0 &
    0.00029 & 0.00033 & 0.00031 & 0.0028 \\
ZnS wurtzite [111] & 4 & 0.141 & 0.152 & 0 & 0 &
    0 & 0 & 0.0038 & 0.017 \\
ZnS wurtzite [221] & 16 & 0.165 & 0.152 & 0 & 0 &
    0.00003 & 0.00002 & 0.0039 & 0.017 \\
\hline
\multicolumn{10}{l}{failed solutions} \\
\hline
CaTiO$_3$ [111] & 20 & 0.4967 & 0.902 & 0.52 & 0.072 &
    0.16 & 0.14 & 0.17 & 1.6 \\
TiO$_2$ rutile [111] & 6 & 0.5358 & 0.758 & 0.40 & 0.009 &
    0.081 & 0.24 & 0.00004 & 0.94 \\
\hline
\multicolumn{10}{p{\textwidth}}{
$C_d$, $C_c$ -- distance and atom overlap cost as defined in equations
(\ref{eq;ligacost}), (\ref{eq;coloringcost})

$s_x$, $s_y$, $s_z$ -- standard deviation in fractional coordinates
normalized to a simple [111] cell

$s_r$ (Å) -- root mean square displacement of the solved sites from the
reference CIF positions
} \\
\hline
\end{tabular}
\end{table}

The structure determination did not work for 2 lower-symmetry
samples of CaTiO$_3$ and TiO$_2$ rutile.  In both of these cases,
the simulated structure showed significantly lower distance cost
$C_d$ while its atom overlap $C_c$ was an order of magnitude higher
than for the correct structure and clearly indicated an unphysical result.
Such results were caused by a poor quality of the extracted
distances, which contained significant errors and omissions with
respect to an ideal distance list.  The peak search and distance extraction
is more difficult for lower symmetry structures, because their pair
distances are more spread and produce small features that can be
below the technique resolution.  Because of poor distance data,
the Liga algorithm converged to incorrect geometries that actually
displayed a better match with the input distances.  Both CaTiO$_3$ and
TiO$_2$ were easily solved when run with ideal distances calculated
from the CIF structure.

The results in Table~\ref{tab;SolvedStructures} suggest several ways
to extend the method and improve its success rate.  First, the
Liga geometry solution and coloring steps can be performed
together, in other words the structure coloring step needs to be merged to
a chemistry aware Liga procedure.  Since atom overlap
cost $C_c$ is meaningful and can be easily evaluated for partial
structures, the total cost minimized by the Liga algorithm should
equal a weighted sum of $C_c$ and distance cost $C_d$.  Such a cost
definition would steer the Liga algorithm away from faulty
structures found for CaTiO$_3$ and TiO$_2$ rutile, because both of them
had huge atom overlaps $C_c$.  Another improvement is to perform PDF
refinement for a full sized structure and update its cost formula so
that the PDF fit residuum $R_w$ is used instead of distance cost $C_d$.
Such modification would prevent the cost advantage for wrong structures
due to errors and omissions in the extracted distances.  The assumption
is that the distance data are still good enough to let
the Liga algorithm construct the correct structure in one of its many
trials.  Finally, the cost definition for partial structures can be
enhanced with other structural criteria such as bond valence sums (BVS)
agreement~\cite{brese;acb91,norbe;jac09}.  Bond valence sums are
not well determined for incomplete intermediate structures and thus
cannot fully match their expected values.  However,
BVS are always increasing, therefore a BVS of some ion that is
significantly larger than its expected value is a clear sign of
such a partial structure's poor quality.

\section{Conclusions}

We have demonstrated the Liga algorithm for structure determination from
PDF can be extended from its original scope of single-element non-periodic
molecules~\cite{juhas;n06,juhas;aca08} to multi-component crystalline
systems.  The procedure assumes known lattice parameters and it solves
structure geometry by optimizing pair distances to match the PDF
extracted values, while the chemical assignment is obtained from
minimization of the atomic radii overlap.  The procedure was tested
on x-ray PDF data from 16 test samples, of which in 14 cases it gave
the correct structure solution.  These are promising results, considering
the technique is at a prototype stage and will be further developed to
improve its ease of use and rate of convergence.  The procedure can be
easily amended by a final PDF refinement step.  Such an implementation
could significantly reduce the overhead in PDF analysis of crystalline materials,
because its most difficult step, a design of suitable structure model,
would become fully automated.


\ack{\textbf{Acknowledgements}}

We gratefully acknowledge Dr.~Emil Božin, Dr.~Ahmad Masadeh and
Dr.~Douglas Robinson for help with x-ray measurements at the
Advanced Photon Source at the Argonne National Laboratory (APS, ANL).
We thank Dr.~Christopher Farrow for helpful suggestions and
consultations and Dr.~Christos Malliakas for providing NaCl and ZnS
wurtzite samples.  We appreciate the computing time and support
at the High Performance Computing Center of the Michigan State
University, where we performed all calculations.  This work has
been supported by the National Science Foundation (NSF) Division of
Materials Research through grant DMR-0520547.  Use of the APS is supported by the U.S. DOE, Office
of Science, Office of Basic Energy Sciences, under Contract
No. W-31-109-Eng-38. The 6ID-D beamline in the MUCAT sector at the APS is
supported by the U.S. DOE, Office of Science, Office of
Basic Energy Sciences, through the Ames Laboratory under
Contract No. W-7405-Eng-82.







\end{document}